\documentstyle[namedreferences,timesv,epsfig]{spackap} 

\def\alwaysmath#1{\ifmmode{#1}\else{$#1$}\fi}

\def\hethree{\ifmmode{\,^{3}{\rm He}^{+}}\else{\thinspace $^{3}{\rm 
He}^{+}$}\fi}
\def\ceta{\ifmmode{\,{\rm C}171\eta}\else{\thinspace ${\rm C}171\eta$}\fi}
\def\heeta{\ifmmode{\,{\rm He}171\eta}\else{\thinspace ${\rm He}171\eta$}\fi}
\def\hxi{\ifmmode{\,{\rm H}213\xi}\else{\thinspace ${\rm H}213\xi$}\fi}
\def\heta{\ifmmode{\,{\rm H}171\eta}\else{\thinspace ${\rm H}171\eta$}\fi}

\def\lesssim{\mathrel{\hbox{\rlap{\hbox{\lower4pt\hbox{$\sim$}}}\hbox{$<$}}}}
\def\gtrsim{\mathrel{\hbox{\rlap{\hbox{\lower4pt\hbox{$\sim$}}}\hbox{$>$}}}}
\let\la=\lesssim			

\begin{opening}
\title{The primordial Helium-4 abundance determination: systematic effects}

\author{Trinh Xuan \surname{Thuan}}
\institute{Dept. of Astronomy, University of Virginia, Charlottesville VA 
22903}
\author{Yuri. I. \surname{Izotov}}
\institute{Main Astronomical Observatory, Kyiv 03680, Ukraine}

\date{}

\end{opening}

\runningauthor{Thuan \& Izotov}
\runningtitle{The primordial abundance of $^4$Helium}

\begin{document}


\begin{abstract}

By extrapolating to O/H = N/H = 0 the empirical correlations $Y$--O/H and $Y$--N/H 
defined by a relatively large sample of $\sim$ 45 Blue Compact Dwarfs (BCDs),
 we have obtained a primordial $^4$Helium mass fraction 
$Y_{\rm p}$= 0.2443$\pm$0.0015 with d$Y$/d$Z$ = 2.4$\pm$1.0. 
This result is in excellent agreement with the average  
$Y_{\rm p}$= 0.2452$\pm$0.0015 determined in the two most metal-deficient BCDs known,
I Zw 18 ($Z_\odot$/50) and SBS 0335--052 ($Z_\odot$/41), where the correction 
for He production is smallest.  The quoted error (1$\sigma$) of 
$\la$ 1\% is statistical and does not include systematic effects.
We examine various systematic effects including collisional excitation 
of Hydrogen lines, ionization structure and temperature fluctuation 
effects, and underlying stellar He {\sc i} absorption, and conclude that 
combining all systematic effects, our $Y_{\rm p}$ may be underestimated 
by $\sim$ 2--4\%.
Taken at face value, our $Y_{\rm p}$ implies a baryon-to-photon
number ratio $\eta$ = (4.7$^{+1.0}_{-0.8}$)$\times$10$^{-10}$ and 
a baryon mass fraction $\Omega_b$$h^2_{100}$ = 
0.017$\pm$0.005 (2$\sigma$), consistent with the values obtained from 
deuterium and Cosmic Microwave Background measurements. Correcting 
$Y_{\rm p}$ upward by 2--4\% would make the agreement even better.

\end{abstract}


%
%

\section{Standard big bang nucleosynthesis}

The standard hot big bang model of nucleosynthesis (SBBN) is one of the 
key quantitative tests of big bang cosmology, along with the 
Hubble expansion and the cosmic microwave background radiation. In the SBBN, 
four light isotopes, D, $^3$He, $^4$He and $^7$Li, were produced by nuclear
reactions a few minutes after the birth of the Universe. 
Given the number of relativistic neutrino species and the
neutron lifetime, the abundances of these light elements depend 
on one cosmological parameter only, the baryon-to-photon ratio $\eta$, 
which in turn is directly related to the density of ordinary baryonic 
matter $\Omega_b$.
The ratio of any two primordial abundances, for example that of D to H 
gives $\eta$, and accurate measurements of the other three light elements,
for example $^4$He/H, tests SBBN.

Of all light elements, the abundance of deuterium (D) is the most sensitive 
to the baryonic density. The primordial D abundance can be 
measured directly in low-metallicity absorption line systems in the spectra 
of high-redshift quasars. The quasar is used as a background light source,
and the nearly primordial gas doing the absorbing is in the outer regions of 
intervening galaxies or in the intergalactic medium (the so-called Lyman 
$\alpha$ clouds).
Tytler and his group (see Tytler et al. 2000 for a review) have 
vigorously 
pursued this type of measurements. They have now obtained D/H measurements 
in the line of sight towards 4 quasars. Combining all measurements, they found
all their data are consistent with a single primordial 
value of the D/H ratio: (D/H)$_{\rm p}$ = 3.0$\pm$0.4$\times$10$^{
-5}$ (O'Meara et al. 2001). This latest value is about 10\% lower than their 
previous value 
(D/H)$_{\rm p}$ = 3.39$\pm$0.25$\times$10$^{-5}$ (Burles \& Tytler 1998).
 
The primordial abundance of $^3$He is also quite sensitive to the 
baryon density, though less than the D abundance. 
It has not been yet measured, mainly because low-mass 
stars make a lot of $^3$He, increasing its value in the interstellar medium 
of the Milky Way well above the primordial value. Furthermore, the amount of 
$^3$He destroyed in stars is unknown. Bania, Rood \& Balser (2000) have 
measured an average $^3$He/H = 1.5$\pm$0.6$\times$10$^{-5}$ in Galactic H 
{\sc ii}
regions. This value represents the average in the interstellar medium of 
the Milky Way, but there exists no good way to extrapolate the $^3$He abundance to the primordial value. 

Old halo stars that formed from nearly pristine gas with very low iron 
abundances during the gravitational collapse of the Milky Way show 
approximately constant $^7$Li/H (the so called ``Spite plateau'', 
Spite \& Spite 
1982), implying that their $^7$Li is nearly primordial. Creation or depletion 
of $^7$Li may make the $^7$Li abundances of halo stars deviate from 
the primordial value. Creation of $^7$Li in the interstellar medium 
by cosmic ray spallation prior to the formation of the Milky Way has to be 
less than 10--20\%, so as not to produce more Be than is observed 
(Ryan, Norris \& Beers 1999).
There is still considerable debate concerning the possible depletion 
of $^7$Li at the surface of stars. 
Depletion mechanisms that have been proposed 
include mixing due to rotation or gravity waves, mass loss in stellar 
winds and gravitational settling. Depending on the exact depletion 
mechanism, the primordial lithium abundance varies 
from ($^7$Li/H)$_{\rm p}$ = (1.73$\pm$0.21)$\times$10$^{-10}$ (Bonifacio 
\& Molaro 1997) to (2.24$\pm$0.57)$\times$10$^{-10}$ (Vauclair \& Charbonnel 
1998), to (3.9$\pm$0.85)$\times$10$^{-10}$ (Pinsonneault et al. 1999).

Because of the relatively large uncertainties in the determination of the 
primordial abundances of $^3$He and $^7$Li, the 
primordial abundance of $^4$He plays a key role for deriving $\Omega_b$
independently of D measurements, and is crucial for checking the consistency 
of SBBN. We discuss next how the primordial $^4$He mass fraction 
 $Y_{\rm p}$ is determined, and the uncertainties 
which enter in such a determination.

\section{The primordial $^4$He abundance as derived from blue compact 
dwarf galaxies}

Because of the relative insensitivity of $^4$He production to the 
baryonic density of matter,  $Y_{\rm p}$ needs to be 
determined to a precision of about one  percent
 to provide useful cosmological 
constraints. This precision can in principle be achieved by 
obtaining high signal-to-noise ratio spectra of a class of star-forming 
dwarf galaxies called Blue Compact Dwarf (BCD) galaxies.
These are low-luminosity ($M_B$ $\geq$ --18) 
systems undergoing an intense burst of star formation in a very 
compact region (less than 1 kpc) which dominates the light of the galaxy 
and which shows blue colors and a H {\sc ii} region-like emission-line 
optical spectrum (Thuan \& Izotov 1998a). 
BCDs are ideal laboratories in which to measure 
the primordial $^4$He abundance because, 
with an oxygen abundance O/H ranging between 1/50 and 1/3 that of the 
Sun, BCDs are among the most metal-deficient gas-rich galaxies known.
Their gas has not been processed through many generations of stars, and thus  best approximates the pristine primordial gas. Izotov \& Thuan (1999) have 
argued that BCDs with O/H less than $\sim$ 1/20 that of the Sun may be 
genuine young galaxies, with ages less than several 100 Myr. 
Thus $Y_{\rm p}$ can be derived accurately in very 
metal-deficient BCDs with only a small correction for Helium made in stars.
Moreover, the theory of nebular emission is well 
understood enough to allow to 
convert He emission-line strengths into abundances with the desired accuracy.

\begin{figure}[hbtp]
    \hspace*{2.0cm}\psfig{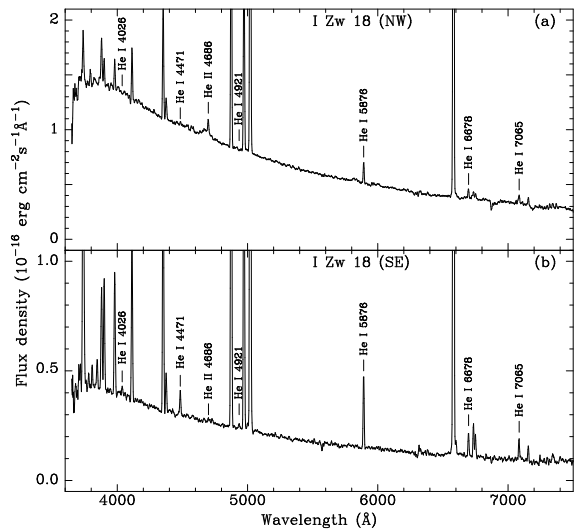}
    \caption{MMT spectra of the NW (top) and SE (bottom) regions of 
I Zw 18. It is evident that underlying He {\sc i} stellar absorption 
is much more important in the NW than in the SE component. All 
marked He {\sc i} lines in the spectrum of the SE component are in emission, while the He {\sc i} 4026 and 4921 lines are in absorption and the He {\sc i} 4471 emission line is barely seen in the spectrum of the NW component.  }
\end{figure}

$Y_{\rm p}$ is generally determined  by 
linear extrapolation of the correlations $Y$--O/H and $Y$--N/H to O/H = N/H = 0 as first proposed by Peimbert \& Torres-Peimbert (1974), 
 where $Y$, N/H and O/H are respectively the 
$^4$He mass fraction, the Oxygen and Nitrogen abundances relative to Hydrogen
of a sample of dwarf irregular and BCD galaxies.
Based on a relatively large sample of $\sim$ 45 BCDs, we 
(Izotov, Thuan \& Lipovetsky 1994, 1997, hereafter ITL94, ITL 97; Izotov \& Thuan 1998, hereafter IT98) have obtained
$Y_{\rm p}$= 0.2443$\pm$0.0015 with d$Y$/d$Z$ = 2.4$\pm$1.0
(see Thuan \& Izotov 1998b, 2000 for reviews). This result is quite 
robust as it is in excellent agreement with the average  
$Y_{\rm p}$= 0.2452$\pm$0.0015 (Izotov 
et al. 1999) determined in the two most metal-deficient BCDs known,
I Zw 18 ($Z_\odot$/50) and SBS 0335--052 ($Z_\odot$/41), where the correction 
for He production is smallest. The quoted statistical error (1$\sigma$) is 
$\la$ 1\% because 
of the high quality of the spectra and the large size of the sample. 

Our derived value of $Y_{\rm p}$ is significantly higher than those 
of 0.228$\pm$0.005 by Pagel et al. (1992), of 0.234$\pm$0.002 
obtained by Olive, Steigman \& Skillman (1997), of 0.2345$\pm$0.0026 by 
Peimbert, Peimbert \& Ruiz (2000) and of 0.2384$\pm$0.0025 by 
Peimbert, Peimbert \& Luridiana (2001). These values are lower than 
ours by several 
times the quoted statistical error and lead to very different cosmological
consequences. This implies that systematic effects in the  
determination of $Y_{\rm p}$ play an important role and need to be considered 
carefully to achieve the desired accuracy. 
We discuss the most important systematic effects below, specifically in  
the two most metal-deficient BCDs known,
I Zw 18 and SBS 0335--052. Because different systematic 
effects are at play in these two objects, they allow to illustrate 
the relative importance of each nicely.  

\section{Systematic effects}

\subsection {Reddening}

One possible systematic effect may be due to the adopted interstellar 
extinction curve. We use the Galactic extinction curve by Whitford (1958).
There is evidence that the extinction curve does change when  
metallicity decreases, being steeper at short wavelengths (e.g. 
Rocca-Volmerange et al. 1981), but the changes are mainly in the ultraviolet 
and are small in the optical range. 
Relying mainly on the three Balmer line ratios H$\alpha$/H$\beta$,
H$\gamma$/H$\beta$ and H$\delta$/H$\beta$, we have adopted an iterative 
procedure to derive simultaneously both the extinction and the absorption 
equivalent width for the hydrogen lines assumed to be 
the same for all lines  (ITL94).
Since the derived extinction correction is applied to the He {\sc i} lines, 
any uncertainty in it will propagate in the derivation of the final He 
abundance. A stringent observational check for the 
adequacy of the extinction curve is the good agreement between the 
corrected intensities of the Balmer hydrogen emission lines and the 
theoretical values for hydrogen recombination line intensities.
Using Monte-Carlo simulations of the hydrogen Balmer ratios, 
Olive \& Skillman (2001) estimate the uncertainties in the reddening corrections to be about 1--2 \% for the blue He lines and 3--4 \% for the red ones.
However, the uncertainties for the red lines
 can be substantially less if the corrected H$\alpha$/H$\beta$ ratio 
matches the theoretical ratio.     

\subsection{Underlying stellar absorption in He {\sc i} lines}

Underlying stellar absorption in He {\sc i} lines 
caused by hot stars can decrease
the intensities of nebular He {\sc i} lines. Model calculations of synthetic 
absorption line strengths in star forming regions by Olofsson (1995) show that
the equivalent widths of He {\sc i} absorption lines decrease as the starburst
ages.
Furthermore, the dependence
of He {\sc i} equivalent widths on metallicity is small, and the 
equivalent width of the
He {\sc i} 4471 absorption line can be as high as 0.35\AA. Unfortunately,
similar calculations for other important lines used in the 
determination of $Y_{\rm p}$ such as He {\sc i} 5876, 
6678 and 7065 are not yet available. 
The effect of underlying He {\sc i} stellar absorption is most important for the
emission lines with the smallest equivalent widths. Therefore, 
the He {\sc i} 5876 emission line which has the largest equivalent width 
is the least affected by such absorption, while the He {\sc i} 4471
emission line is the most affected because of its low equivalent width, 
 the effect of underlying absorption being 5--10 times larger for it than for 
the 5876 line.

\begin{figure}
    \hspace*{2.0cm}\psfig{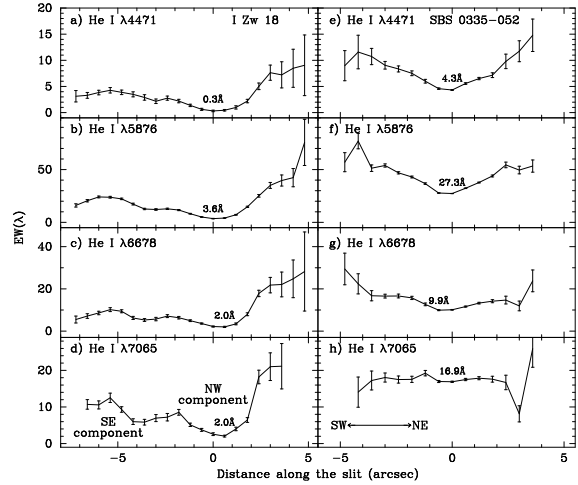}
    \caption{Spatial distributions of the He {\sc i} nebular emission 
line equivalent widths in I Zw 18 (left panel) and in SBS 0335--052 (right 
panel). The error bars are 1$\sigma$ deviations. The value of the minimum 
equivalent width for each He {\sc i} emission line is given.}
\end{figure}

Figure 1 shows the spectra of the 2 brightest centers of star formation in 
I Zw 18, the NW and SE components, with the He {\sc i} lines marked. 
It is clear that the NW component suffers 
far more stellar absorption than the SE one: all marked He {\sc i} lines 
of the SE component are in emission while the two He {\sc i} 4026 and 
4921 lines are in absorption and the He {\sc i} 4471 line is hardly 
detected in the NW component, its intensity being decreased by 
a factor of $\sim$ 2 (Izotov et al. 1999).
 However, the strong He {\sc i} absorption in the
NW component of I Zw 18 constitutes the exception rather than the rule. 
It is relatively less
important in other very metal-deficient BCDs, such as SBS 0335--052, 
because the equivalent widths of their
He {\sc i} emission lines are considerably larger than those of the
NW component of I Zw 18. This can be seen in Figure 2 which compares the 
spatial distributions of the He {\sc i} nebular emission line 
equivalent widths ($EW$) in I Zw 18 and SBS 0335--052. 
While the maximum $EW$s of the He I emission lines
in the SE component of I Zw 18 are close to those in the central brightest 
part of SBS 0335--052, the He {\sc i} line $EW$s in the NW component of
I Zw 18 are several times smaller. The largest ratio of minimum
values of equivalent widths in I Zw 18 as compared to in SBS 0335--052 
is $\sim$ 14 for the 4471 line.
 The effect of underlying stellar absorption is smaller for the other 
He {\sc i} lines because of their higher equivalent widths. 
Thus for the 6678 line, Izotov et al. (1999) estimate it to be $\sim$ 5\% 
in the NW component of 
I Zw 18, but to be less than 1\% in its SE component and in SBS 0335--052.
As for the 5876 line, it is less than 0.4\% in SBS 0335--052 (it is 
contaminated by the Galactic Na {\sc i} 5890 line in I Zw 18, so is unusable).
It was mainly because the importance of underlying He {\sc i} 
stellar absorption 
was insufficiently recognized in the NW component of I Zw 18 that led to the 
low $Y_{\rm p}$ values of Pagel et al. (1992) and Olive et al. (1997).
\begin{figure}[hbtp]
    \hspace*{4.0cm}\psfig{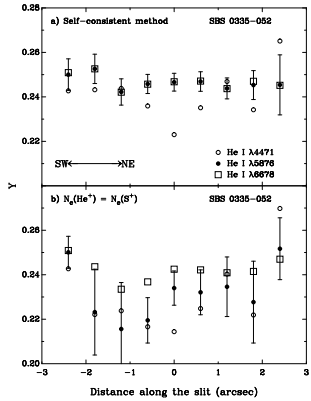}
    \caption{The spatial distributions of the helium mass fractions
in SBS 0335--052 derived from the He {\sc i} 4471, 5876
and 6678 emission line intensities. The intensities of the He {\sc i}
emission lines in a) are corrected for fluorescent and collisional
enhancement with an electron number density $N_e$(He {\sc ii}) and an optical 
depth $\tau$(3889) derived self-consistently from the observed 
He {\sc i} 3889, 4471, 5876, 6678 and 7065 emission line intensities. 
The points for the 4471 line are below the points for the other lines 
because of underlying  He {\sc i} stellar absorption. 
For comparison, the intensities of the He {\sc i} emission
lines in b) are corrected only for collisional enhancement, with an
electron number density $N_e$(S {\sc ii}). 
The points for the different lines do not agree anymore.
The 1$\sigma$ error bars are shown only for 
the He mass fraction derived from the He {\sc i} $\lambda$5876 emission line. 
They are 
larger in b) because of the large uncertainties in the determination of 
$N_e$(S {\sc ii}).}
\end{figure}

In our work thus far, we have not corrected self-consistently for underlying 
He {\sc i} self-absorption. We have simply not used regions where underlying 
absorption is important (such as the NW component in I Zw 18), or not 
averaged in lines that give a $Y$ clearly deviant from the $Y$s from other lines.
Thus, in the case of SBS 0335--052, Figure 3a shows that the $Y$ obtained from
He {\sc i} 4471 line is systematically below the values from the 
He {\sc i} 5876 and 6678 lines. It is affected by underlying absorption 
and is not included. In the future, we plan to solve for He {\sc i} 
absorption 
(assumed to be the same for all He {\sc i} lines) 
self-consistently, as we did for the H {\sc i} lines (see below). 

\subsection{He {\sc i} and H {\sc i} emissivities}

Line ratios corrected for reddening and absorption are converted to 
He/H abundance ratios by using theoretical emissivities calculated 
from recombination and radiative cascade theory. We use the H {\sc i} 
emissivities of Brocklehurst (1971) which are in excellent 
agreement with the more recent ones by Hummer \& Storey (1987) for the 
range of temperatures and densities in metal-deficient BCDs, and the 
He {\sc i} emissivities of Smits (1996). Benjamin, Skillman \& Smits (1999) 
have 
estimated that uncertainties in the theoretical He {\sc i} emissivities 
can be as large as 1.5\%, $\sim$ 3 times worse than the accuracy expected 
from comparing Brocklehurst (1972) and Smits (1996) emissivities.
      
\subsection{Electron density determination}
 
To determine element abundances, we adopt a two-zone photoionized H II
region model (ITL94,ITL97,IT98): 
a high-ionization zone with temperature $T_e$(O {\sc iii}), where the 
O {\sc iii}, Ne {\sc iii} and Ar {\sc iv} lines originate, and a 
low-ionization zone with
temperature $T_e$(O {\sc ii}), where the 
O {\sc ii}, N {\sc ii}, S {\sc ii} and Fe {\sc iii} lines originate. 
As for the Ar {\sc iii} and S {\sc iii} lines they originate  in the intermediate zone
between the high and low-ionization regions.
The [S {\sc ii}] $\lambda$6717/$\lambda$6731 line ratio is used to determine the
electron density $N_e$(S {\sc ii}) according to five-level atom 
calculations.

Previous authors have set  $N_e$(He {\sc ii}) =  $N_e$(S {\sc ii}). However, 
this is not appropriate as $N_e$(S {\sc ii}) measures the density in the 
low-ionization zone, while He {\sc ii} is produced in the high-ionization 
zone. Furthermore, the [S {\sc ii}] ratio is fairly insensitive for 
densities below 100 cm$^{-3}$.  
Thus it is much better to determine  $N_e$(He {\sc ii}) directly from the 
He {\sc i} lines themselves, in a self-consistent manner, which we have done. Because the electron density enters linearily in the calculation of the 
correction for collisional enhancement of the He {\sc i} emission lines 
(see below), its different estimate by us and previous authors 
is also responsible for the difference between our $Y_{\rm p}$ and theirs.   

\begin{figure}[hbtp]
    \hspace*{2.0cm}\psfig{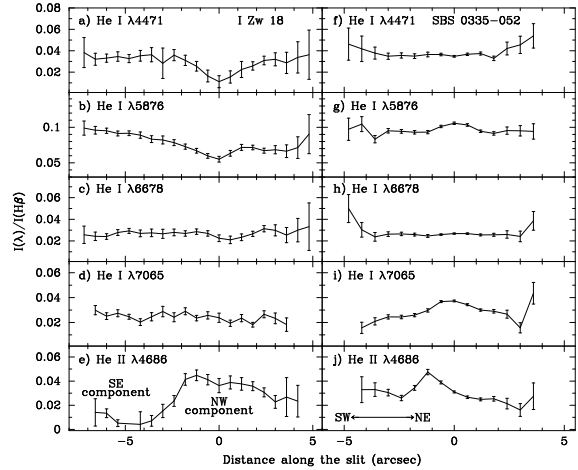}
    \caption{The spatial distributions of the He {\sc i} and He {\sc ii} 
4686 nebular emission line intensities in I Zw 18 
(left panel) and in SBS 0335--052 (right panel). The error bars are 1$\sigma$ 
deviations.}
\end{figure}

\subsection{The collisional excitation of hydrogen lines}    

It is also generally assumed that case B recombination theory holds,
i.e that the line photons (usually resonance lines of abundant ions) are scattered so many times that their downward radiative transitions can effectively 
be neglected. In other words, the H {\sc ii} regions are considered to be 
optically thick in the Lyman transitions but optically thin in the other 
transitions of both hydrogen and helium atoms. (The opposite assumption,
case A, where all emitted photons escape without absorption and there is no radiative transfer problem, does not apply to nebulae with large enough 
amounts of gas and optical depths to be seen). 
However, there are known physical effects that make the helium lines 
deviate from case B. One such effect is 
the collisional excitation of the Balmer hydrogen lines by 
thermal electrons.

This process and its effect  
on the determination of $Y_{\rm p}$ was first discussed by 
Davidson and Kinman (1985). Low-metallicity H {\sc ii} regions have high 
enough electron temperatures so that the observed fluxes of lines like 
H$\alpha$ may be overestimated by this effect, leading to  
an underestimate of $Y_p$. 
Davidson \& Kinman (1985) estimated crudely this effect to be 
$\sim$ 2\% for I Zw 18. Sasselov \& Goldwirth (1995) examined 
this effect with detailed radiative transfer calculations and also found 
it to cause an increase of $\sim$ 2\% in the $Y$ of I Zw 18 and up to 3\% for 
other metal-deficient BCDs. Stasi\'nska \& Izotov (2001) used a grid of 
photoionization models to show that the effect of collisional excitation 
on the H$\alpha$/H$\beta$ ratio can be as high as 8\%, resulting in an 
upward correction in $Y$ of up to 5\% in objects like SBS 0335--052.
Thus this effect can be one of the most important sources of 
systematics in the determination of $Y_{\rm p}$.

\subsection{The collisional and fluorescent enhancements of 
He {\sc i} lines}

In the high range of electron temperatures found in metal-deficient BCDs
(10000--20000 K), collisional excitation from the metastable level 2 $^3$S 
level of He {\sc i} can be important in populating the higher levels and 
making the He {\sc i} line intensities deviate from pure recombination 
values.
Another effect that also leads to deviation from pure recombination values is 
self-absorption in some optically thick emission lines 
which populates the upper 
levels of He {\sc i}, a mechanism called fluorescence. The emission lines 
most sensitive to fluorescence in the optical range are the     
 He {\sc i} 3889 and 7065 lines. The He {\sc i} 7065 also  
plays an important role because it is particularly density sensitive.
In contrast to collisional excitation,
which increases the line intensities of all He {\sc i} lines, the 
fluorescent mechanism decreases the intensity of the  He {\sc i} 3889 
line as its optical depth increases, while increasing the intensities 
of the He {\sc i} 4471, 5876, 6678, and 7065 lines.

\begin{figure}[hbtp]
    \hspace*{3.0cm}\psfig{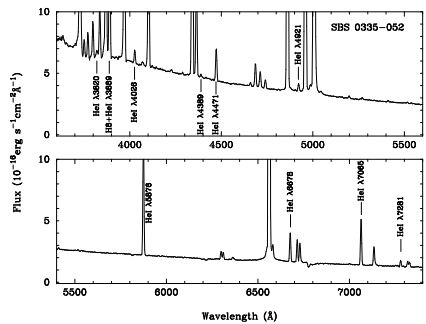}
    \caption{Keck spectrum of the BCD SBS 0335--052 with labeled He {\sc i}
emission lines.}
\end{figure}

How do these two effects play out in I Zw 18 and SBS 0335--052? 
They are less important in I Zw 18 than in SBS 0335--052 because       
the physical conditions in the two BCDs are quite different. 
While in I Zw 18 the electron number density is 
small ($N_e$(S {\sc ii}) $\leq$ 100 cm$^{-3}$) and collisional enhancement 
has a minor effect on the derived helium abundance, 
the electron number density in SBS 0335--052
is considerably higher ($N_e$(S {\sc ii}) $\sim$ 500 cm$^{-3}$ in the central 
part of the H {\sc ii} 
region). Additionally, the linear size of the H {\sc ii} 
region in SBS 0335--052 is $\sim$ 5 times larger than in I Zw 18, 
making it optically thick for some He {\sc i} transitions. 
Therefore, both collisional and 
fluorescent enhancements of He {\sc i} emission lines 
play a significant role in this galaxy (Izotov et al. 1999). 
Figure 4 shows that the 
spatial distribution of He {\sc i} emission line intensities 
in SBS 0335--052 is very different from that in I Zw 18. 
The increase of He {\sc i} 5876 and 
He {\sc i} 7065 emission line strengths by $\sim$ 20\% and $\sim$ 75\% 
respectively
in the central part of SBS 0335--052 within a radius $\sim$ 2\arcsec\ is
caused by collisional and fluorescent enhancement. The increase
of the He~{\sc i} $\lambda$6678 emission line intensity is only $\leq$ 4\%.
The combined effect of collisional enhancement and underlying stellar absorption
results in a small depression in the He {\sc i} 4471 
intensity in the central region. 
As for I Zw 18, the main effect is underlying He {\sc i} stellar absorption 
causing the 
dip in the He {\sc i} emission line intensities at the location of the 
NW component. 

To correct the  He {\sc i} line intensities for 
collisional enhancement, we use the correction factors calculated 
by Kingdon \& Ferland (1995) based on collisional rates 
by Sawey \& Berrington (1993). To correct the 
 He {\sc i} line intensities for fluorescent enhancement, 
we have fitted the Robbins (1968) correction factors with polynomials 
as given in IT98.
Since the collisional enhancement factor of the He {\sc i}
lines depend exponentially on the electron temperature and linearly on the 
electron density $N_e$ (He {\sc ii}), we correct for these two effects 
and determine 
 $N_e$ (He {\sc ii}) at the same time,
 in a self-consistent way so that the He {\sc i} 
5876/4471, 6678/4471, and 7065/4471 line ratios have their recombination 
values, after correction for both collisional and 
fluorescent enhancement.

\subsection{The non-coincidence of the H$^+$ and He$^+$ 
Str\"omgren spheres}

Depending on the hardness of the ionizing 
radiation, the radius of the  He$^+$ 
sphere can be smaller or larger than the radius of the  H$^+$ sphere.
When the ionizing radiation is soft, the first case prevails, 
it is necessary to consider the presence of unseen neutral helium in 
the  H$^+$ zone and a correction needs to be made,
resulting in a higher He abundance. On the other hand, if 
the ionizing radiation is hard, the second case holds, 
there is neutral hydrogen in the He$^+$ zone, which 
results in a downward correction of the     
 He abundance. The correction in $Y$ can be as 
high as several percent, either upward or downward 
(e.g. Steigman, Viegas \& Gruenwald 1997, 
Peimbert, Peimbert \& Ruiz 2000, Sauer \& Jedamzik 2001).
The hardness of the radiation is usually characterized by the
``radiation softness parameter'' $\eta$ defined by V\'ilchez \& Pagel 
(1988) as $ \eta = \frac{{\rm O}^+}{{\rm S}^+}\frac{{\rm S}^{++}}{{\rm O}^{++}}$, and the ionization parameter $U$.
For both I Zw 18 and SBS 0335--052, the ionizing radiation is hard,
and the correction to $Y$ is downward.  
Using the extensive grid of correction factors as functions 
of $\eta$ and $U$,
calculated by Sauer \& Jedamzik (2001) 
using photoionized H {\sc ii} region models, we found the downward 
correction to be less than 1\% for both BCDs.

\subsection{Temperature structure}

To convert the He {\sc i} line intensities into abundances, we have set 
the electron temperature $T_e$(He {\sc ii}) to be equal to  $T_e$(O {\sc iii})
as obtained from the [O {\sc iii}] 
4363 / (4959+5007) ratio. However as 
emphasized by Peimbert, Peimbert \& Ruiz (2000), 
detailed modeling of BCDs like I Zw 18 
(Stasi\'nska \& Schaerer 1999) and examination of 
photoionization models (Stasi\'nska 1990) 
suggest that  $T_e$(He {\sc ii}) is smaller than  $T_e$(O {\sc iii}) by at 
least 5\% in this type of object (deviations from a constant temperature  
are sometime called  ``temperature fluctuations''). 
Peimbert, Peimbert \& Luridiana (2001) found that this leads to a 
downward correction of Y of about 3\%.
However, examination of the correction factors for 
temperature effects derived by Sauer \& Jedamzik (2001) 
using photoionization models give downward corrections of Y that are 
considerably smaller for I Zw 18 and SBS 0335--052, less than 1\%.   
   
\section{Cosmological implications}

In summary, the important systematic effects that may affect our 
determination of $Y_{\rm p}$ from spectra of the two most metal-deficient 
BCDs known, I Zw 18 (the SE component)
 and SBS 0335--052, and that have not been taken into 
account by our self-consistent procedure are: 
1) the collisional excitation of 
Hydrogen lines that can increase $Y_{\rm p}$ by up to 5\%; 2) 
the non-coincidence of the H$^+$ and He$^+$ zones that may decrease $Y_p$
by $\sim$ 1\%; 3) the temperature fluctuations that may decrease it 
by  $\sim$ 1 -- 3\%; and 4) underlying stellar He {\sc i} absorption 
that may 
increase it by $\sim$ 1\%. Thus, combining all those systematic effects, 
our $Y_{\rm p}$ value may be underestimated by as much as $\sim$ 2 -- 4\%.       

Taken at face value, our $Y_{\rm p}$ = 0.2452$\pm$0.0015 implies a baryon-to-photon
number ratio $\eta$ = (4.7$^{+1.0}_{-0.8}$)$\times$10$^{-10}$.
 This translates to a baryon mass fraction $\Omega_b$$h^2_{100}$ = 
0.017$\pm$0.005 (2$\sigma$)  where $h_{100}$ 
is the Hubble constant in units of 100 km s$^{-1}$Mpc$^{-1}$.
This value is consistent with the one of 0.020$\pm$0.002 (2$\sigma$) 
derived from the primordial deuterium abundance measured in high-redshift 
hydrogen clouds backlit by distant quasars (O'Meara et al. 2001),
vindicating SBBN.
It is also consistent with the baryon mass fraction $\Omega_b$$h^2_{100}$ = 
0.022$\pm$0.003 (1$\sigma$)
inferred from measurements of the angular power spectrum 
of the Cosmic Microwave Background (CMB, Netterfield et al. 2001). Note that 
correcting  upward $Y_{\rm p}$ by 2--4\% would bring it into even better 
agreement with the deuterium and CMB measurements. 

\section{Future work} 

It is clear from the previous discussion that, to decrease or eliminate 
the main systematic effects in the determination of $Y_{\rm p}$, 
it is best to determine all the  
four following quantities -- the electron 
density $N_e$ (He {\sc ii}) and temperature $T_e$ (He {\sc ii}) of the He {\sc ii} zone, the optical depth $\tau$ (3889) in the He {\sc i} 3889 line, 
and the equivalent 
width for underlying He {\sc i} absorption -- in a totally self-consistent 
manner.
ITL94, ITL97 and IT98 have used the five He {\sc i}  
3889, 4471, 5876, 6678 and 7065 lines to solve self-consistently 
for $N_e$ (He {\sc ii}) and $\tau$ (3889). We plan to add the He {\sc i} 
4026, 4438 and 4922 lines, giving a total of 8 lines, 
to solve self-consistently for
all four quantities. The last three lines are particularly sensitive to 
underlying He {\sc i} stellar absorption. However their intensities are 
less than 1/10 
that of the 5876 line, and very high signal-to-noise ratio spectra are 
needed to determine their intensities precisely. 
Such a spectrum obtained with the Keck telescope and 
where all 8 He {\sc i} lines are marked  
is shown in Figure 4 for the BCD SBS 0335--052.

Future work to improve on the determination of $Y_{\rm p}$
 will consist of: 1) obtaining deep spectra for the 
most metal-deficient BCDs known, to determine 
 $Y_{\rm p}$ using a self-consistent 
method based on 8 lines; 2) detailed modeling of BCDs such as done 
for I Zw 18 by Stasi\'nska \& Schaerer (1999). It was found that 
a simple photoionization model is insufficient to account for the 
high [O {\sc iii}] 4363/5007 ratio in I Zw 18 
and that an additional heating mechanism 
such as shock heating is necessary. Such a modeling will also help 
to quantify potential temperature fluctuations;  and 3) searching
 for more extremely metal-deficient BCDs such as I Zw 18 and SBS 0335--052, 
to increase the number of objects where we can determine $Y_{\rm p}$ without 
a large correction for He production by stars. \\\\

We thank Dr. Johannes Geiss for a careful reading of the manuscript. 
This work has been supported by National Science Foundation grant 
AST-9616863.

\end{document}